\documentclass{article}
\usepackage{spconf,amsmath,amsfonts,graphicx,mathrsfs}
\usepackage{cite}
\usepackage{enumerate}
\newcommand{\minimize}{\mathop{\rm minimize}\limits}

\newcommand{\argmin}{\mathop{\rm argmin}\limits}
\newcommand{\prox}{\mathrm{prox}}
\def\inner<#1>{\langle #1 \rangle}

\title{Designing nearly tight window for improving time-frequency masking}
\name{Tsubasa Kusano, Yoshiki Masuyama, Kohei Yatabe and Yasuhiro Oikawa}
\address{Department of Intermedia Art and Science, Waseda University, Tokyo, Japan}
\begin{document}
\ninept
\maketitle
\begin{abstract}
\vspace{-2pt}
Many audio signal processing methods are formulated in the time-frequency (T-F) domain which is obtained by the short-time Fourier transform (STFT).
The properties of the STFT are fully characterized by window function, number of frequency channels, and time-shift. Thus, designing a better window is important for improving the performance of the processing especially when a less redundant T-F representation is desirable.
While many window functions have been proposed in the literature, they are designed to have a good frequency response for analysis, which may not perform well in terms of signal processing.
The window design must take the effect of the reconstruction (from the T-F domain into the time domain) into account for improving the performance.
In this paper, an optimization-based design method of a nearly tight window is proposed to obtain a window performing well for the T-F domain signal processing.
\end{abstract}
\vspace{-4pt}
\section{Introduction}
\vspace{-4pt}
\label{sec:intro}

Many audio signal processing methods are formulated as modifications of the signal in the time-frequency (T-F) domain, which is often called T-F masking.
For converting the signal into the T-F domain, the short-time Fourier transform (STFT) \cite{Gabor1946}%
\footnote{
STFT is also often called the Gabor transform based on \cite{Gabor1946}.
Note that some literature strictly distinguishes STFT from the Gabor transform by their mapping properties \cite{Analysis1998}, while others do not.
In this paper, we may utilize the term ``STFT'' in the sense of the discrete Gabor transform (DGT), which is a common habit especially in the acoustical signal processing community.
}
is usually utilized owing to its simplicity and easily understandable structure \cite{Walnut1992,Analysis1998,Søndergaard2012,Moreno-Picot2018,Ephraim1984,Yilmaz2004,Gerkmann2012}.
While most of the research has concentrated on the method of modification in the T-F domain (the way how to construct a T-F mask), the method of converting a signal into the T-F domain is also important for improving the performance of processing.

When STFT is considered as the conversion to the T-F domain, its property is fully characterized by the window function since STFT is a highly structured transform.
Aiming to obtain a better T-F representation, many window functions have been proposed to improve their frequency responses \cite{Slepian1961,Kaiser1980,Harris1978,Nuttall1981,Adams1991,Yiu2013,Kawahara2017}.
For example, the Hann window is one popular window which has a good sidelobe decay.
The Nuttall window was proposed to achieve a better sidelobe decay, while the Kaiser window was proposed so that its frequency response was adjustable by a tuning parameter.
Although such research on window functions has provided a better T-F representation, most research only has considered the \textit{analysis} side.
That is, there is little research on window functions considering the \textit{reconstruction} point of view.

To realize processing in the T-F domain, the signal must be reconstructed back into the time domain after the T-F-domain processing.
The reconstruction side of STFT is achieved by the (pseudo-) inverse STFT which also involves a window function.
Therefore, for the T-F domain signal processing, a window function must be chosen in accordance with not only STFT but also the inverse STFT.
Indeed, incorrect choice of the pair of window functions (for STFT and the inverse STFT) makes reconstruction impossible.
To allow a reasonable window for the reconstruction, the T-F representation is usually chosen to be \textit{redundant}, and the error-minimizing window called the canonical dual window is often used for the reconstruction (see Section~\ref{sec:preliminaries}).

For some applications favoring less redundant T-F representation, choice of the window function is more critical for the reconstruction (and thus, critical for processing).
One example is T-F masking in low-power devices which allow a little computation \cite{Jeub2012,Parchami2016}.
In such cases, redundancy should be lowered because higher redundancy directly results in higher computational cost.
Another very important example is speech enhancement based on deep learning.
Recent study has shown that non-redundant T-F representation can improve the performance of enhancement using deep neural networks (DNN) \cite{Koizumi2018}.
This is because less redundant T-F representation reduces the number of parameters to be learned, which makes the training easier.
For those applications, redundancy of STFT should be lowered by increasing the window shifting width.
However, the inverse transform becomes more sensitive to the error of signal processing when the redundancy is reduced (see Section~\ref{subsec:influence_of_condition_number}), which also degrades the performance.
Although there exists a type of window insensitive to such processing error, called \textit{tight} window, it has a drawback that its frequency response is often poor (sidelobe level is high).
Therefore, a window function which is less sensitive to processing error and, at the same time, has a good frequency response is desired for realizing a better processing in a less redundant situation.

In this paper, we propose a window design method to simultaneously meet both requirements.
It aims to make a window function closer to a tight window, while its frequency response is constrained to be better.
Since the designed window is not strictly tight, we call it \textit{nearly tight} window.
The proposed method is formulated as an optimization problem so that it can easily control the trade-off between the two requirements, and it is solved by the linearized alternating direction method of multipliers (ADMM).

\vspace{-4pt}
\section{Preliminaries}
\vspace{-4pt}

\label{sec:preliminaries}

While the discrete and downsampled T-F transform is called ``STFT'' in acoustical signal processing, the literature of T-F analysis calls it the discrete Gabor transform (DGT) \cite{Analysis1998}.
Hereafter, we will utilize the language used in DGT to express the T-F representation because it will be easier for explaining the proposed method.

\vspace{-4pt}
\subsection{Gabor system and discrete Gabor transform (DGT)}
\vspace{-2pt}

Let a window be denoted by $\mathbf{g} = [\mathbf{g}[0],\mathbf{g}[1],\ldots,\mathbf{g}[L-1]]^T\!\in \mathbb{R}^L$.
DGT is a T-F transform based on a collection of windowed sinusoids,
\begin{equation}
\mathcal{G}({\mathbf g}, a, M) = \left\{ {\mathbf g}_{m,n} \right\}_{m=0,\ldots, M-1, \;n=0, \ldots, N-1},
\label{eq:GabSyst}
\end{equation}
which is called the Gabor system, where $a\in \mathbb{N}$ is the time-shifting width, $M\in \mathbb{N}$ is the number of frequency channels,
\begin{equation}
  {\mathbf g}_{m,n}[l] = \mathrm{e}^{\mathrm{i}\frac{2\pi ml}{M}}{\mathbf g}[l-an],
\end{equation}
is a windowed complex sinusoid, and $\mathrm{i} = \sqrt{-1}$.
DGT of a discrete signal ${\mathbf f} \in \mathbb{R}^L$ is defined by the following inner product:
\vspace{-4pt}
\begin{equation}
  (\mathbf{G}_\mathbf{g}\mathbf{f})[m+nM] = \inner<\mathbf{f}, {\mathbf g}_{m,n}> = \sum_{l=0}^{L-1} \mathbf{f}[l]\,\overline{{\mathbf g}_{m,n}[l]}.
  \vspace{-4pt}
\end{equation}
where $\overline{x}$ is the complex conjugate of $x$, and $\mathbf{G}_\mathbf{g}\in\mathbb{C}^{MN\times L}$ is the matrix consisting of all the elements in the Gabor system in Eq.~\eqref{eq:GabSyst}.
That is, multiplying $\mathbf{G}_\mathbf{g}$ to a signal obtains the vectorized version of its T-F representation which is often called ``spectrogram.''

\vspace{-4pt}
\subsection{Reconstruction of time-domain signal from T-F domain}
\vspace{-2pt}

A system $\mathcal{G}({\mathbf g}, a, M)$ is said to be a \textit{frame} \cite{Daubechies1992,christensen2003} if there exist $0<A, B <\infty$ such that
\begin{equation}
  A\left\Vert {\mathbf f} \right\Vert_2^2 \le \sum_{m,n} \vert \inner<{\mathbf f},{\mathbf g}_{m,n}> \vert^2 \le B \left\Vert {\mathbf f} \right\Vert_2^2,
  \vspace{-4pt}
\end{equation}
for all ${\mathbf f} \in \mathbb{R}^L$, where $\Vert \cdot \Vert_p$ is the $\ell_p$ norm. $A$ and $B$ are called the lower and upper frame bound, respectively.
If the Gabor system is a frame, a time-domain signal can be reconstructed from its T-F domain representation.
The inverse DGT, reconstructing a signal from its coefficients $\mathbf{c} \in \mathbb{C}^{MN}$, with respect to $\mathcal{G}(\mathbf{g}, a, M)$ is defined by
\begin{equation}
  {\mathbf f}_\text{syn} = \sum_{n,m} \mathbf{c}[m+nM] \,\mathbf{g}_{m,n} = \mathbf{G}_\mathbf{g}^* \mathbf{c},
  \vspace{-4pt}
\end{equation}
where $\mathbf{G}_\mathbf{g}^*$ denotes the complex-conjugate transpose of $\mathbf{G}_\mathbf{g}$.
If a Gabor system $\mathcal{G}(\mathbf{g}, a, M)$ is a frame, then there exists the corresponding dual Gabor frame $\mathcal{G}(\mathbf{h}, a, M)=\{\mathbf{h}_{m,n} \}$ which satisfies
\begin{equation}
  {\mathbf f} = \sum_{n,m} \inner<\mathbf{f}, {\mathbf g}_{m,n}> \mathbf{h}_{m,n},
  \label{eq:reconst}
  \vspace{-4pt}
\end{equation}
where $\mathbf{h}_{m,n}[l] = \mathrm{e}^{\mathrm{i}\frac{2\pi ml}{M}}{\mathbf h}[l-an]$, and $\mathbf{h}$ is a dual window of $\mathbf{g}$.
That is, a time-domain signal can be reconstructed if (1) $\mathcal{G}(\mathbf{g}, a, M)$ is a frame, and (2) $\mathbf{h}$ is a dual window of $\mathbf{g}$.
These conditions are decided by the window pair $\mathbf{g},\mathbf{h}$, the time-shifting width $a$, and the number of frequency channels $M$.

When a Gabor system $\mathcal{G}(\mathbf{g}, a, M)$ is redundant, the corresponding dual window $\mathbf{h}$ is not unique, and infinitely many variation of $\mathbf{h}$ can satisfy the reconstruction formula, Eq.~\eqref{eq:reconst}.
One standard choice among all possible dual windows is the canonical dual window
\begin{equation}
  \tilde{\mathbf g} = \mathbf{S}_\mathbf{g}^{-1}\mathbf{g},
\end{equation}
where $\mathbf{S}_\mathbf{g} = \mathbf{G}_\mathbf{g}^* \mathbf{G}_\mathbf{g}$ is the so-called frame operator defined as
\begin{equation}
  \mathbf{S}_\mathbf{g}\mathbf{f} = \sum_{m,n}  \inner<\mathbf{f}, {\mathbf g}_{m,n}> \mathbf{g}_{m,n} = (\mathbf{G}_\mathbf{g}^* \mathbf{G}_\mathbf{g}) \,\mathbf{f}.
  \vspace{-4pt}
\end{equation}
The canonical dual window is optimal in the sense that its synthesis operator corresponds to the Moore--Penrose pseudo-inverse:
\begin{align}
  \sum_{n,m} \mathbf{c}[m+nM] \,\tilde{\mathbf g}_{m,n} &= \sum_{n,m} \mathbf{c}[m+nM] \,\mathbf{S}_{\mathbf g}^{-1}{\mathbf g}_{m,n} \nonumber \\[-2pt]
  &= (\mathbf{G}_\mathbf{g}^* \mathbf{G}_\mathbf{g})^{-1} \mathbf{G}_\mathbf{g}^* \mathbf{c}.
  \label{eq:pseudo-inverse}
\end{align}
In this paper, the canonical dual window is considered for inverse DGT as it is the standard choice in acoustical signal processing.
One reason for such popularity should be because of the optimality to the following least squares signal reconstruction problem:
\begin{equation}
  \minimize_\mathbf{x} \,\,\,\, \left\Vert \mathbf{G}_\mathbf{g} \mathbf{x} - \hat{\mathbf{c}} \right\Vert_2^2,
  \vspace{-2pt}
\end{equation}
whose solution is $\mathbf{G}_{\tilde{\mathbf g}}^*\hat{\mathbf{c}}$ as can be confirmed from the fact in Eq.~\eqref{eq:pseudo-inverse}.

\begin{figure}[t]
	\centering
	\includegraphics[width=0.99\columnwidth]{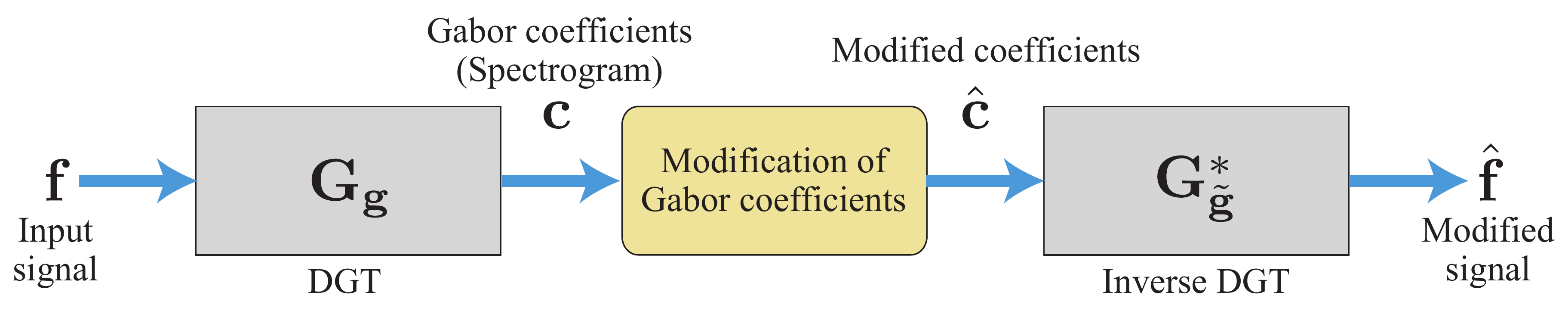}
   \vspace{-10pt}
	\caption{Framework of the signal processing in the T-F domain.}
	\label{fig:diagram}
\end{figure}

\begin{figure}[t]
	\centering
	\includegraphics[width=0.99\columnwidth]{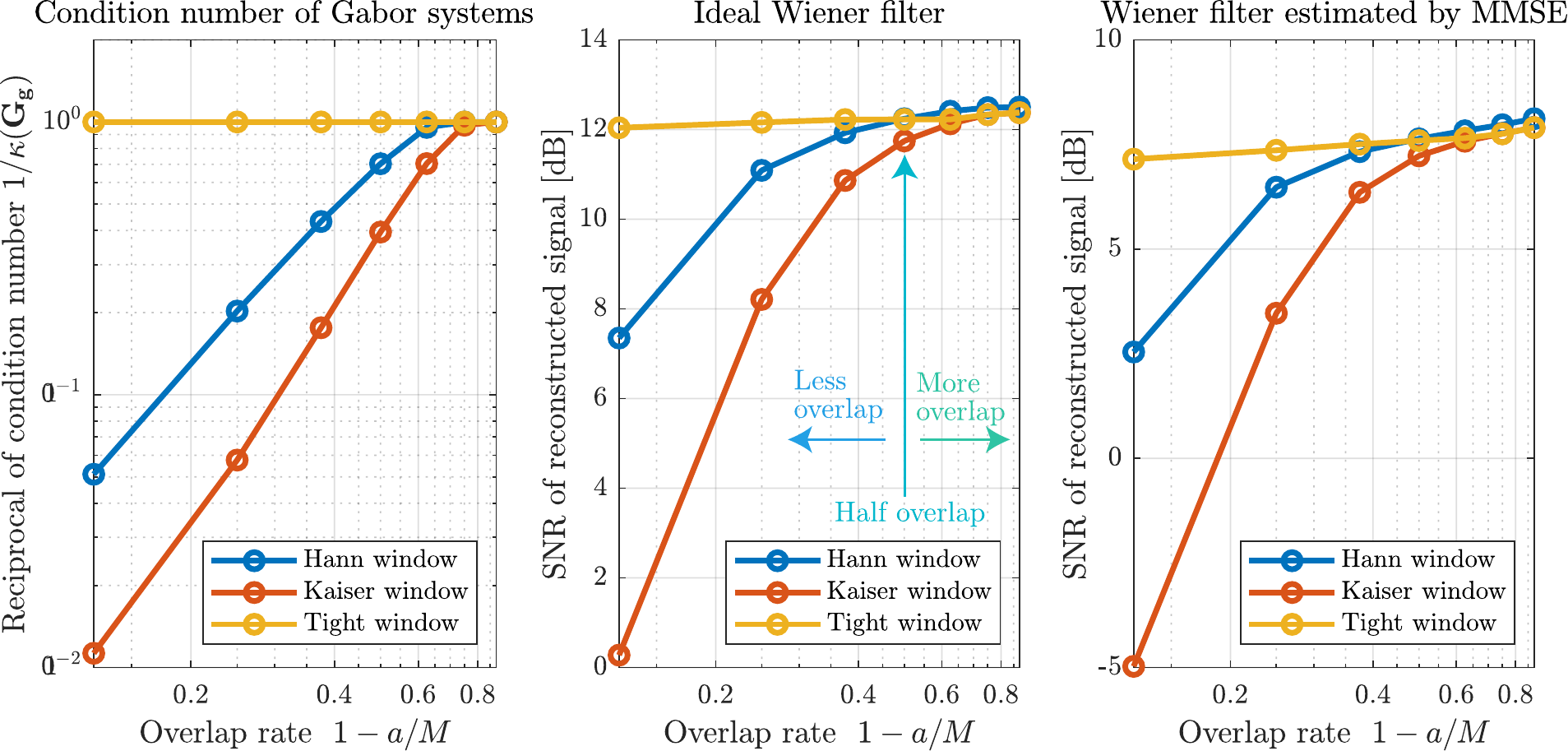}
	\vspace{-8pt}
	\caption{(left) Condition number of the DGT matrix $\kappa(\mathbf{G}_\mathbf{g})$. (center) Denoising result using the ideal Wiener filter. (right) Denoising result using the Wiener filter with MMSE noise power estimation.}
	\label{fig:influence_of_cn}
\end{figure}

\vspace{-4pt}
\subsection{Influence of window functions on signal processing}
\vspace{-2pt}
\label{subsec:influence_of_condition_number}

A signal processing framework in T-F domain is illustrated in Fig~\ref{fig:diagram}.
In words, some processing is performed in the T-F domain to modify the Gabor coefficient $\mathbf{c}$ to $\hat{\mathbf c}$, and then the inverse DGT is applied to obtain the processed result $\hat{\mathbf{f}}$.
While the quality of the processing is important for obtaining a good result, the transformation pair, DGT and inverse DGT, is also important since it decides the coefficient $\mathbf{c}$.

To see the effect of the window pair in terms of T-F domain signal processing, a preliminary experiment was performed.
200 speech signals \cite{Mowlaee2016} from TIMIT database \cite{TIMIT} were degraded by adding the Gaussian noise in the time domain so that the signal-to-noise ratio (SNR) became 0 dB.
They were enhanced by the Wiener filter (T-F masking based on the power ratio of noisy and clean signals) with a minimum mean-square error (MMSE) estimator of noise power \cite{Gerkmann2012} and the decision-directed approach \cite{Ephraim1984}.
The redundancy was changed by changing $a$ while fixing the window length to 256 and $M=256$.
Its performance was compared with the ideal Wiener filter and the condition number of $\mathbf{G}_\mathbf{g}$,
\begin{equation}
  \kappa(\mathbf{G}_\mathbf{g}) = \sigma_{\mathrm{max}}(\mathbf{G}_\mathbf{g})\,/\,\sigma_{\mathrm{min}}({\mathbf{G}_\mathbf{g}})= \sqrt{B\,/A\,},
  \label{eq:condition_number}
\end{equation}
which is the standard measure of numerical stability of Eq.~\eqref{eq:pseudo-inverse}, where $\sigma_{\mathrm{max}}(\mathbf{G}_\mathbf{g})$ and $\sigma_{\mathrm{min}}({\mathbf{G}_\mathbf{g}})$ denote the maximum and minimum singular value of $\mathbf{G}_\mathbf{g}$, respectively.

Three types of window functions were utilized for comparison: the Hann window, the Kaiser window $(\alpha = 10)$, and the canonical tight window of the Kaiser window.
A window $\mathbf{g}_\text{T}$ is said to be \textit{tight} if its canonical dual window is itself (i.e., self-dual) \cite{Cvetkovic1998a}.
Then,
\begin{equation}
  \mathbf{S} = \mathbf{G}_{\mathbf{g}_\text{T}}^* \mathbf{G}_{\mathbf{g}_\text{T}} = A\,\mathbf{I}
\end{equation}
holds, where $\mathbf{I}$ is the identity.
Thus, the condition number of a tight window is always 1.
Particularly, a tight window with $A = 1$ is called the Parseval tight window.
The canonical tight window of a window $\mathbf{g}$ can be obtained by inverting square root of the frame operator:
\begin{equation}
  \mathbf{g}_\text{T} = \mathbf{S}_\mathbf{g}^{-\frac{1}{2}}\mathbf{g},
  \label{eq:canonical_tight}
\end{equation}
which corresponds to the solution of the following problem \cite{Janssen2002}:
\begin{equation}
  \minimize_\mathbf{x \in \mathcal{T}} \,\,\,\, \left\Vert \mathbf{g} - \mathbf{x} \right\Vert_2,
  \vspace{-2pt}
  \label{eq:canonical_tight2}
\end{equation}
where $\mathcal{T}$ is the set of all Parseval tight windows.
Thus, the canonical tight window is the closest Parseval tight window from the window $\mathbf{g}$.
An efficient algorithm for its computation is available in the LTFAT toolbox \cite{Søndergaard2012a}, see also \cite{Janssen2002}.

Results of the experiment are shown in Fig.~\ref{fig:influence_of_cn}, where SNR is the average among all speech signals.
For both ideal and realistic Wiener filters (center and right), the processing performances of the Hann and Kaiser windows were degraded as the redundancy decreased (horizontal axes are related to the redundancy).
In contrast, the performance of the tight Kaiser window was not degraded much.
These results can be predicted from the condition numbers (left).
Based on this experiment, a window $\mathbf{g}$ should be designed so that the condition number $\kappa(\mathbf{G}_\mathbf{g})$ becomes lower.
A tight window is the best window in this sense because its condition number is the lowest.
However, as in the figure, a tight window is not always the best in terms of processing, which is because the frequency response of a tight window is usually not better than one of a non-tight window (see Fig.~\ref{fig:windows}).
Therefore, in this paper, a design method of a \textit{nearly tight} window is proposed so that the condition number is lowered while its frequency response is kept well.

\vspace{-4pt}
\subsection{Related works on Gabor window design}
\vspace{-2pt}
\label{subsec:related}

For designing a low-condition-numbered window, design methods of tight windows have been proposed \cite{Cvetkovic2000,Perraudin2018}.
These methods aim to find a tight window with better frequency responses.
However, since the constraint to the tight window greatly limits the set of variables, desired characteristics may not be obtained.

On the other hand, some methods of nearly-tight window design have been proposed \cite{Second-order2004,Jiang2013,Wilbur2004}.
One approach of this research is to minimize the difference between the frame operator and identity operator using the gradient-based optimization \cite{Second-order2004,Jiang2013}.
These methods minimize the distance to the set of tight windows by the gradient method, whereby they have a possibility of falling into the local minima.
Another approach is to replace the non-convex cost of measuring the distance to the tight window with convex functions \cite{Wilbur2004}.
Since that method is formulated as convex optimization, it is guaranteed that globally optimal solutions can be obtained, so a trade-off between the condition number and the frequency response can be easily considered.
However, as a result of approximating the cost function, the obtained solutions may not be close to the original solution which is tight. The cost should be reduced strictly without approximation, while the trade-off should be easily adjusted.

\vspace{-4pt}
\section{Proposed method}
\vspace{-2pt}
\label{sec:proposed}

In this section, we propose a design method of nearly tight window that can easily control the trade-off between the desired frequency response and the condition number.
At first, we formulate the nearly tight Gabor window design as a constrained minimization problem.
Then, an algorithm solving this problem through the proximal operators is introduced.
Since a window whose support is shorter than the signal length is used in most signal processing, the formulation considers $\mathbf{g}[l] = 0$ for $l=K,\ldots, L-1$, i.e., only $\mathbf{g}[l]$ $(l=0,\ldots,K-1)$ are treated as the variables in this paper.

\vspace{-4pt}
\subsection{Problem formulation for designing nearly tight window}
\vspace{-2pt}

To propose an easily adjustable window function design, the desired frequency response is considered as a constraint, and the window is made closer to a tight window as possible.
Its direct formulation is
\begin{equation}
  \minimize_{\mathbf{g}\in \mathcal{C}}\,\,\,\, \frac{1}{2}d_\mathcal{T}^2(\mathbf{g}),
\end{equation}
where $d_\mathcal{T}(\mathbf{g})$ is the distance to the set of Parseval tight windows $\mathcal{T}$,
\begin{equation}
  d_\mathcal{T}(\mathbf{g}) = \min_{\mathbf{x} \in \mathcal{T}} \Vert \mathbf{g} -\mathbf{x} \Vert_2,
\end{equation}
and $\mathcal{C}$ is the set of windows satisfying the desired frequency response.
Since the magnitude response should be considered in decibels for audio applications, a popular choice for the set $\mathcal{C}$ to constrain the frequency response into desired one, in filter design \cite{Wu1999}, is
\begin{equation}
  \mathcal{C} = \{\mathbf{g} \in \mathbb{R}^K \mid \Vert \log_{10}\vert \tilde{\mathbf{F}} \mathbf{g} \vert - \log_{10}\mathbf{d} \Vert_\infty \le \log_{10}\beta \},
  \label{eq:constraint_logChebyshev}
\end{equation}
where $\mathbf{d}\in \mathbb{R}_+^{\tilde K}$ ($\tilde{K}>K$) is magnitude of the desired frequency response, $\tilde{\mathbf{F}} \in \mathbb{C}^{\tilde{K} \times K}$ is the zero-padded discrete Fourier transform,
\begin{equation}
  \tilde{\mathbf{F}}[m,n] = \frac{1}{\sqrt{\tilde K}}\mathrm{e}^{-\mathrm{i}\frac{2 \pi mn}{\tilde K}},
\end{equation}
and $\beta \ge 1$ is a parameter for controlling the amount of error.
However, directly treating this constraint is not easy because taking difference after absolute value results in the non-convex set.

Since the requirement in window design (in contrast to filter design) is to lower the sidelobe level towards zero (i.e., increasing the magnitude of sidelobe is usually not desired), it should be sufficient to constrain only the upper bound.
Based on this observation,
\begin{equation}
  \tilde{\mathcal{C}} = \{\mathbf{g} \in \mathbb{R}^K \mid \vert (\tilde{\mathbf{F}}\mathbf{g})[n] \vert \le \beta \mathbf{d}[n]\,\,\,\text{for}\,\, n = 0,\ldots,N-1 \},
  \label{eq:convexConstSet}
\end{equation}
is considered as the constraint set instead of Eq.~\eqref{eq:constraint_logChebyshev}.
Consequently, our formulation becomes a minimization problem on the convex set:
\begin{equation}
  \minimize_{\mathbf{g} \in \tilde{\mathcal{C}}}\,\,\,\, \frac{1}{2} d^2_\mathcal{T}(\mathbf{g}).
  \label{eq:optimization_problem}
\end{equation}
This model directly handles the distance function $d_\mathcal{T}$ instead of approximation as in \cite{Wilbur2004}, while the desired frequency response is strictly imposed by the constraint $\tilde{\mathcal{C}}$ as opposed to \cite{Second-order2004,Jiang2013}.

\begin{figure*}[t]
	\centering
	\includegraphics[width=1.98\columnwidth]{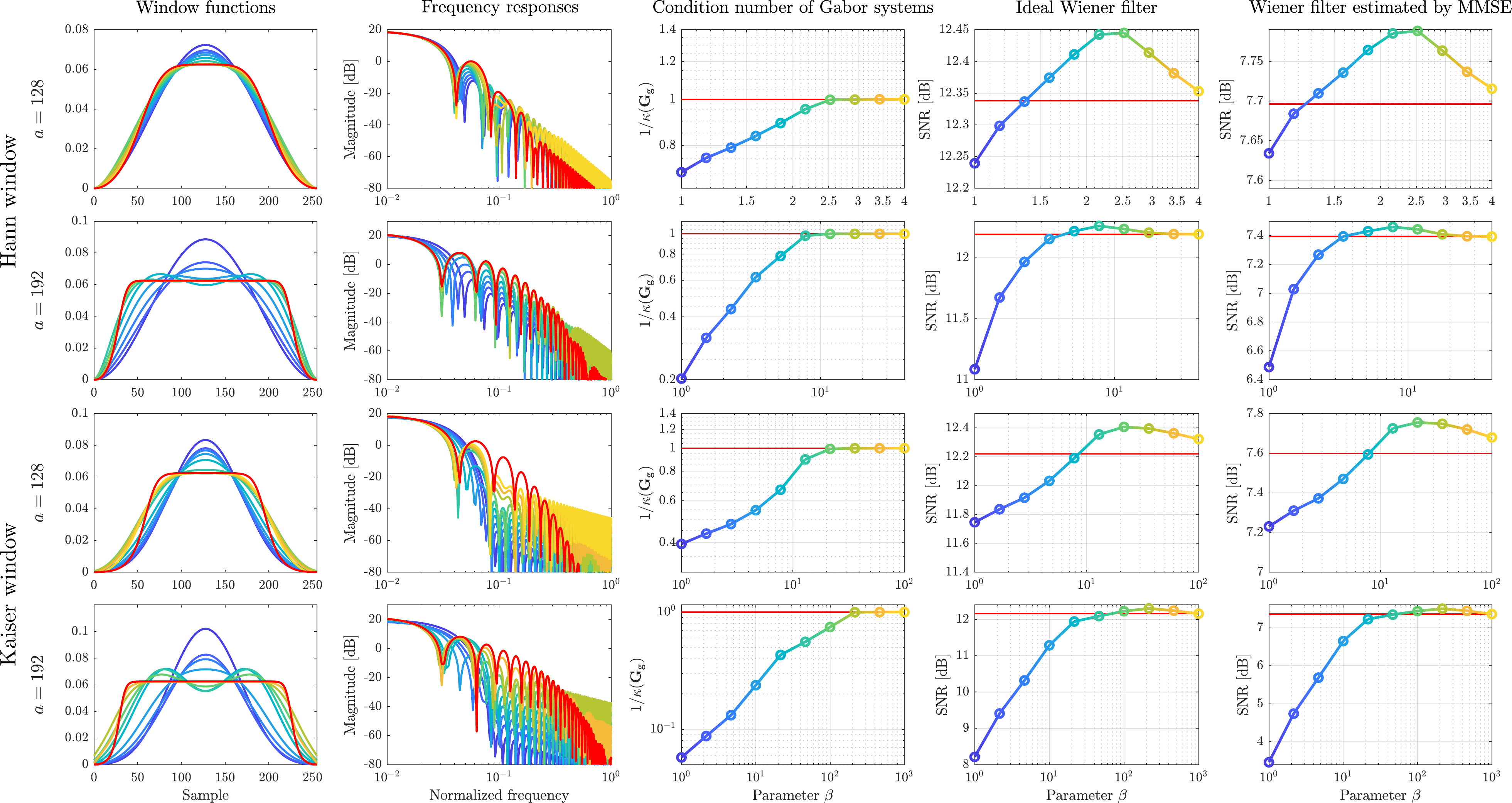}
      \vspace{-8pt}
	\caption{Designed nearly tight windows by the proposed method.
	Each column shows (from left to right) the obtained window shapes, their frequency responses, their condition numbers, denoising results for the ideal Wiener filter, and those for the Wiener filter with MMSE noise power estimation.
	Each row shows (from top to bottom) the results of the Hann-based windows for $a=128$, $192$, and those of the Kaiser-based windows ($\alpha=10$) for $a=128$, $192$.
	The transition of colors from blue to yellow represents a change in parameter $\beta$ of the proposed method, where the blue represents $\mathbf{g}^\mathrm{o}$ and brighter color (larger $\beta$) means closer to tight.
	Red lines indicate the canonical tight window of $\mathbf{g}^\mathrm{o}$.}
	\label{fig:windows}
\end{figure*}

\vspace{-4pt}
\subsection{Algorithm for solving problem using linearized ADMM}
\vspace{-2pt}

To solve Eq.~\eqref{eq:optimization_problem}, linearized ADMM \cite{Zhang2010,Yang2012,Parikh2013} is utilized in this paper.
It is an algorithm solving problems written in the following form:
\begin{equation}
  \minimize_\mathbf{x}\,\,\, \mathscr{F}(\mathbf{x}) + \mathscr{G}(\mathbf{A}\mathbf{x}),
  \label{eq:lADMM}
\end{equation}
where $\mathscr{F}(x)$ and $\mathscr{G}(x)$ are proper lower semi-continuous functions, and $\mathbf{A}$ is a linear operator.
By using the proximity operator \cite{Parikh2013},
\begin{equation}
  \prox_{\rho \mathscr{F}}(\mathbf{x}) = \argmin_\mathbf{y} \left\{\mathscr{F}(\mathbf{y}) + \frac{1}{2 \rho}\Vert \mathbf{y} - \mathbf{x} \Vert_2^2 \right\},
\end{equation}
the linearized ADMM algorithm is given as the following procedure:
  \begin{align}
    \mathbf{x}^{[k+1]} &= \prox_{\mu \mathscr{F}}\Bigl(\mathbf{x}^{[k]} - \frac{\mu}{\lambda} \mathbf{A}^*(\mathbf{A}\mathbf{x}^{[k]} - \mathbf{z}^{[k]} + \mathbf{u}^{[k]}) \Bigr), \\
    \mathbf{z}^{[k+1]} &= \prox_{\lambda \mathscr{G}}(\mathbf{A}\mathbf{x}^{[k+1]} + \mathbf{u}^{[k]}), \\
    \mathbf{u}^{[k+1]} &= \mathbf{u}^{[k]} + \mathbf{A}\mathbf{x}^{[k+1]} - \mathbf{z}^{[k+1]},
  \end{align}
where $\lambda$ and $\mu$ are real numbers satisfying $0 < \mu \le \lambda /\Vert \mathbf{A} \Vert_\text{op}^2$, and $\Vert \cdot \Vert_\text{op}$ is the operator norm.

For applying this linearized ADMM algorithm to Eq.~\eqref{eq:optimization_problem}, it is rewritten as the equivalent problem having the form of Eq.~\eqref{eq:lADMM}:
\begin{equation}
  \minimize_\mathbf{g} \,\,\, \frac{1}{2} d_\mathcal{T}^2(\mathbf{g}) + \iota(\tilde{\mathbf{F}} \mathbf{g}),
  \label{eq:problemADMMform}
\end{equation}
where $\iota(\mathbf{z})$ is the indicator function corresponding to Eq.~\eqref{eq:convexConstSet},
\begin{align}
  \iota(\mathbf{z}) &= \begin{cases}
    \,\:0 & (\vert \mathbf{z}[n] \vert \le \beta\mathbf{d}[n]\,\,\,\text{for}\,\, n = 0,\ldots,N-1) \\
    \,\infty & (\text{otherwise})
  \end{cases}.
\end{align}
Then, Eq.~\eqref{eq:problemADMMform} is solved by iterating the following procedure:
\begin{align}
\mathbf{g}^{[k+1]} &= \prox_{\frac{\mu}{2} d_\mathcal{T}^2}\left(\mathbf{g}^{[k]} - \frac{\mu}{\lambda} \tilde{\mathbf F}^*(\tilde{\mathbf F}{\mathbf g}^{[k]} - \mathbf{z}^{[k]} + \mathbf{u}^{[k]}) \right), \label{eq:admm1} \\
\mathbf{z}^{[k+1]} &= \prox_{\iota}(\tilde{\mathbf F}{\mathbf g}^{[k+1]}+\mathbf{u}^{[k]}), \label{eq:admm2} \\
\mathbf{u}^{[k+1]} &= \mathbf{u}^{[k]} + \tilde{\mathbf F}{\mathbf g}^{[k+1]} - \mathbf{z}^{[k+1]},
\end{align}
where $\mathrm{prox}_{\frac{\mu}{2} d^2_\mathcal{T}}(\cdot)$ and $\prox_{\iota}(\cdot)$ in Eqs.~\eqref{eq:admm1} and \eqref{eq:admm2} are given by
\vspace{-4pt}
\begin{align}
  \mathrm{prox}_{\frac{\mu}{2} d^2_\mathcal{T}}(\mathbf{g}) &= \frac{1}{1+\mu}\mathbf{g} + \frac{\mu}{1+\mu}\mathbf{S}_\mathbf{g}^{-\frac{1}{2}}\mathbf{g}, \label{eq:prox_d}\\
  \mathrm{prox}_{{\iota}}({\mathbf z})[n] &= \min \left\{ \beta\frac{\mathbf{d}[n]}{\vert \mathbf{z}[n]\vert}, 1\right\}\mathbf{z}[n].
\end{align}
Thanks to the property of the canonical tight window in Eq.~\eqref{eq:canonical_tight2}, Eq.~\eqref{eq:prox_d} can be expected to give an appropriate descent direction even though the cost function $d^2_\mathcal{T}$ is non-convex.
Therefore, this algorithm should be able to effectively manage the difficulty associated with the non-convexity of $d^2_\mathcal{T}$.

\vspace{-4pt}
\section{NUMERICAL EXPERIMENTS}
\vspace{-2pt}
\label{sec:experiment}

The shapes, frequency responses and condition numbers of the windows designed by the proposed method were compared with the denoising performance provided by the same experiment in Section~\ref{subsec:influence_of_condition_number} using ideal and MMSE Wiener filters.
For the initial window inputted to the algorithm, the Hann and Kaiser windows, whose energies were normalized to $a/M$, were chosen in accordance with Section~\ref{subsec:influence_of_condition_number}.
By iterating the algorithm from these windows denoted by $\mathbf{g}^\mathrm{o}$, the designed windows are expected to have characteristics similar to $\mathbf{g}^\mathrm{o}$ with a better condition number.
The frequency responses $\mathbf{d}$ for the constraint set $\tilde{\mathcal{C}}$ were constructed by interpolating the maxima of $\log_{10}\!\vert \tilde{\mathbf{F}} \mathbf{g}^\mathrm{o} \vert$ by the cubic $C^2$-splines.

The obtained nearly tight windows by the proposed method and the denoising results for $a=128$, $192$ are summarized in Fig.~\ref{fig:windows}.
When the parameter $\beta$ was set to a higher value (brighter color), then the obtained windows got closer to a tight window, which can be confirmed by the condition numbers.
Note that the canonical tight window has the highest level of the first side lobe which may prevent a denoising method to be work correctly.
It can be seen that some windows obtained by the proposed method outperformed both the original window (blue) and the canonical tight window (red) in terms of the denoising results.
These results indicate that the proposed method can design a window having better characteristics for T-F domain signal processing than the original and the canonical tight window.
The performance was adjustable by the single parameter $\beta$, which enables to look for a better window by a simple line search.

\vspace{-4pt}
\section{CONCLUSION}
\vspace{-4pt}
\label{sec:conclusion}

In this paper, the nearly tight window designing method for signal processing in the T-F domain is proposed.
The proposed method can obtain nearly tight windows having desired frequency responses, which can result in a better performance of T-F masking than those of original and canonical tight windows.
Future work includes the automatic adjustment of $\beta$ as well as the generalization of the method.


\end{document}